# An Adaptive and Explicit Fourth Order Runge-Kutta-Fehlberg Method Coupled with Compact Finite Differencing for Pricing American Put Options.

Chinonso I. Nwankwo[a,*], Weizhong Dai[b]

[a] Department of Mathematics, Statistics, and Computer Science, University of Illinois at Chicago, Chicago, IL 60607, USA

[b] Department of Mathematics and Statistics, Louisiana Tech University, Ruston LA 71272, USA
[*] Corresponding author, cnwank4@uic.edu, nonsonwankwo@gmail.com

## Abstract

We propose an adaptive and explicit fourth-order Runge-Kutta-Fehlberg method coupled with a fourth-order compact scheme to solve the American put options problem. First, the free boundary problem is converted into a system of partial differential equations with a fixed domain by using logarithm transformation and taking additional derivatives. With the addition of an intermediate function with a fixed free boundary, a quadratic formula is derived to compute the velocity of the optimal exercise boundary analytically. Furthermore, we implement an extrapolation method to ensure that at least, a third-order accuracy in space is maintained at the boundary point when computing the optimal exercise boundary from its derivative. As such, it enables us to employ fourth-order spatial and temporal discretization with Dirichlet boundary conditions for obtaining the numerical solution of the asset option, option Greeks, and the optimal exercise boundary. The advantage of the Runge-Kutta-Fehlberg method is based on error control and the adjustment of the time step to maintain the error at a certain threshold. By comparing with some existing methods in the numerical experiment, it shows that the present method has a better performance in terms of computational speed and provides a more accurate solution.

## 1. Introduction

American style option, written on an asset $S_t$ with the strike price, $K$ and expiration time $T$ differs from the European option due to the early (optimal) exercise boundary which leads to a free boundary problem Let $V(S,\tau)$ denote the option price, $s_f(\tau)$ represent the optimal exercise boundary and $\tau = T - t$. Then, $V(S,\tau)$ satisfies the coupled free boundary value problem:

$$-\frac{\partial V(S,\tau)}{\partial \tau} + \frac{1}{2}\sigma^2 S^2 \frac{\partial^2 V(S,\tau)}{\partial S^2} + rS\frac{\partial V(S,\tau)}{\partial S} - rV(S,\tau) = 0, \quad \text{for } S > s_f(\tau), \tag{1a}$$

$$V(S,\tau) = K - S, \quad \text{for } S < s_f(\tau). \tag{1b}$$

Here, the initial and boundary conditions are given as:

$$V(S,0) = max(K - S, 0), \quad s_f(0) = K; \tag{1c}$$

$$V(s_f, \tau) = K - s_f(\tau), \quad V(0,\tau) = K, \quad V(\infty, \tau) = 0, \quad \frac{\partial}{\partial S}V(s_f, \tau) = -1, \tag{1d}$$

This early exercise boundary presents an advantage and a challenge in the valuation of American options and solving (1), respectively. In terms of advantage, it provides a possibility to exercise the options early, however, we have some level of complexity. This is because the early exercise boundary and the American option values are simultaneously obtained (McKean, 1965) when solving the free boundary problem. It is well known that due to this complexity, there is no closed-form or analytical formula for evaluation of the American option. Hence, numerical, semi-numerical, and analytical approximation present a choice for solving (1).

Several numerical methods have been proposed for solving the American options problem with the front-fixing approach. In particular, the second-order explicit and implicit finite difference schemes have been used (Company et al., 2014; Company et al., 2016; Nielsen et al., 2001; Wu and Kwok, 1997; Ballestra, 2018). Moreover, the degeneracy that occurs in the front-fixing method of Wu and Kwok (1998) was further pointed out in the work of Kim et al. (2013). The finite element method has also been implemented for solving American options based on the front-fixing approach (Holmes and Yang, 2008; Zhang et al., 2014; Song et al. 2017). Holmes and Yang (2008) implemented the Crank-Nicholson method and Zhang et al. (2014) and Song et al. (2017) used the first-order backward method based on the temporal discretization.

Over the decades, embedded high order Runge-Kutta adaptive methods have been developed (Fehlberg, 1969; Verner, 1978, Dormand and Prince, 1980; Cash and Karp, 1990; Papageorgiou and Tsitouras, 1996; Tsitouras, 1998; Macdougall and Verner, 2002; Verner, 2010) and implemented in several works of literature (Simos, 1993; Simos and Papakaliatakis, 1998, Burden et al., 2016) for solving heat problems, stochastic wave and Schrodinger equations (Wilkie and Cetinbas 2005; Treblay and Carrighton, 2004), harmonic oscillator problem (Hoover et al. 2016), thin-film model (Romeo et al. 2008) and Lotka Voltera prey-predator model (Paul et al., 2016). The adaptive Runge-Kutta method, which is more effective than the classical fourth Runge-Kutta method, is based on the control of error estimation which results in time step adjustment and optimal selection of time step at each time level. The variation in time step by optimal selection of the time step at each time level provides some computational benefits. In terms of cost, it enables the selection of large time steps in a local region with sufficient smoothness and small-time steps in a local region with large variation and discontinuity (William and Saul, 1992).

The motivation of this work is to implement an adaptive and explicit fourth-order Runge-Kutta-Fehlberg time integration method coupled with a fourth-order compact scheme for solving the American put options problem based on the front-fixing approach. To the best of our knowledge, we are the first to implement this combination for solving the American option problem. In short, by implementing a method of extrapolation, we improve on the techniques of Kim et al. (2013) which avoid the degeneracy that occurs near the optimal exercise boundary to obtain the velocity of the latter analytically with high order accuracy in space. We then apply adaptive fourth-order Runge-Kutta-Fehlberg methods to compute the optimal exercise boundary with high order accuracy in time. Coupled with a compact finite difference scheme for spatial discretization, more accurate numerical solutions of the option price, and option Greeks are obtained with fast computation.

The rest of the paper is organized as follows. In section 2, we discuss the various transformation involved in our method. In section 3, we employ a compact scheme in the spatial discretization and adaptive Runge-Kutta-Fehlberg method for temporal discretization. In section 4, we investigate and compare the numerical performance of an adaptive Runge-Kutta-Fehlberg with the classical Runge-Kutta and other existing methods and conclude the paper in section 5.

## 2. Transformations and Free Boundary Analysis

### 2.1. Front-Fixing Logarithmic Transformation

Here, we first employ logarithmic transformation (Egorova et. al., 2016; Wu and Kwok, 1997) to fix the free boundary by using the following relation

$$x = \ln \frac{S}{s_f(\tau)} = \ln S - \ln s_f(\tau), \qquad U(x,\tau) = V(S,\tau). \tag{2}$$

Applying it to equation (1), we then have

$$\frac{\partial U(x,\tau)}{\partial \tau} - \frac{1}{2}\sigma^2 \frac{\partial^2 U(x,\tau)}{\partial x^2} - \left(\frac{s'_f}{s_f} + r - \frac{\sigma^2}{2}\right)\frac{\partial U(x,\tau)}{\partial x} + rU(x,\tau) = 0, \qquad x > 0; \tag{3a}$$

$$U(x,\tau) = K - S = K - s_f(\tau)e^x, \qquad \text{for } x < 0; \tag{3b}$$

where the initial condition (1) is changed to

$$U(x,0) = \max(K - Ke^x, 0) = 0, \qquad x \geq 0, \qquad s_f(0) = K. \tag{3c}$$

By letting $x \to 0^-$, we obtain from (3c) that $U(0,\tau) = K - s_f(\tau)$. Thus, together with (1d), we obtain the boundary condition for (3a) as

$$U(0,\tau) = K - s_f(\tau), \qquad U(\infty,\tau) = 0. \tag{3d}$$

Taking further derivative to remove the first-order derivative in (3a), we obtain a system of coupled partial differential equations consisting of the asset, delta, gamma, and speed options as follows:

$$\frac{\partial U(x,\tau)}{\partial \tau} - \frac{1}{2}\sigma^2 \frac{\partial^2 U(x,\tau)}{\partial x^2} - \left(\frac{s'_f(\tau)}{s_f(\tau)} + r - \frac{\sigma^2}{2}\right) W(x,\tau) + rU(x,\tau) = 0, \tag{4a}$$

$$\frac{\partial W(x,\tau)}{\partial \tau} - \frac{1}{2}\sigma^2 \frac{\partial^2 W(x,\tau)}{\partial x^2} - \left(\frac{s'_f(\tau)}{s_f(\tau)} + r - \frac{\sigma^2}{2}\right)\frac{\partial^2 U(x,\tau)}{\partial x^2} + rW(x,\tau) = 0, \tag{4b}$$

$$\frac{\partial Y(x,\tau)}{\partial \tau} - \frac{1}{2}\sigma^2 \frac{\partial^2 Y(x,\tau)}{\partial x^2} - \left(\frac{s'_f(\tau)}{s_f(\tau)} + r - \frac{\sigma^2}{2}\right)\frac{\partial^2 W(x,\tau)}{\partial x^2} + rY(x,\tau) = 0, \tag{4c}$$

$$\frac{\partial Z(x,\tau)}{\partial \tau} - \frac{1}{2}\sigma^2 \frac{\partial^2 Z(x,\tau)}{\partial x^2} - \left(\frac{s'_f(\tau)}{s_f(\tau)} + r - \frac{\sigma^2}{2}\right)\frac{\partial^2 Y(x,\tau)}{\partial x^2} + rZ(x,\tau) = 0, \tag{4d}$$

where $x > 0$, and the initial and boundary conditions for $U(x,\tau), W(x,\tau) Y(x,\tau)$, and $Z(x,\tau)$ are given as:

$$U(x,0) = 0, \quad W(x,0) = 0, \quad Y(x,0) = 0, \quad Z(x,0) = 0, \quad s_f(0) = K, \quad x \geq 0; \tag{4e}$$

$$U(0,\tau) = K - s_f(\tau), \quad W(0,\tau) = Y(0,\tau) = Z(0,\tau) = -s_f(\tau); \tag{4f}$$

$$U(\infty,\tau) = 0, \quad W(\infty,\tau) = 0, \ Y(\infty,\tau) = 0, \quad Z(\infty,\tau) = 0. \tag{4g}$$

## 2.2. Transformed Function with the Fixed Free Boundary

Due to degeneracy that occurs near the optimal exercise boundary, we adopt the idea in the work of Kim et al. (2013, 2017) that implements an intermediate (square root) function to avoid such degeneracy. The transformed function is of the form

$$Q(x,\tau) = \sqrt{U(x,\tau) - K + e^x s_f(\tau)}, \qquad U(x,\tau) = Q^2(x,\tau) + K - e^x s_f(\tau), \tag{5a}$$

with

$$Q(x,\tau) \begin{cases} = 0, & x \in [\ln s_f(\infty) - \ln s_f(0)], \\ > 0, & x \in (0, \infty). \end{cases} \tag{5b}$$

Here, $s_f(\infty)$ is the asymptotically optimal exercise boundary given as follows ():

$$s_f(\infty) = \frac{\gamma}{\gamma + 1} K, \qquad \gamma = \frac{2r}{\sigma^2}. \tag{5c}$$

By computing the higher derivatives of $Q(x, \tau)$ and $U(x, \tau)$ when $x = 0$ using (1) and (5), Kim et al. (2013) obtained the derivative of the optimal exercise boundary by taking the Taylor expansion of $Q(x, \tau)$ near the optimal exercise boundary with up to fourth-order accuracy as follows:

$$Q(\bar{x}, \tau) = Q(0, \tau) + \bar{x} Q'(0, \tau) + \frac{\bar{x}^2}{2} Q''(0, \tau) + \frac{\bar{x}^3}{6} Q'''(0, \tau) + O(\bar{x}^4). \tag{6}$$

Here,

$$Q(0, \tau) = 0, \qquad Q'(0, \tau) = \frac{\varphi}{\sigma}, \qquad Q''(0, \tau) = -\frac{2\xi_\tau \varphi}{3\sigma^3}; \tag{7a}$$

$$Q'''(0, \tau) = \frac{2\xi_\tau^2 \varphi}{3\sigma^5} + \frac{r\varphi}{2\sigma^3}. \tag{7b}$$

Note that

$$\xi_\tau = v + \frac{1}{s_f} \frac{\partial s_f}{\partial \tau}, \qquad v = r - \frac{\sigma^2}{2}, \qquad \varphi = \sqrt{rK}, \tag{7c}$$

and $\bar{x} \ll x$ is arbitrary and very close to the optimal exercise boundary. Substituting (7a)-(7c) into (6), we obtain a quadratic equation with respect to $\partial s_f / \partial \tau$. Thus, the velocity of the optimal exercise boundary is presented in quadratic form as follows:

$$\frac{\partial s_f}{\partial \tau} = \frac{-b - \sqrt{b^2 - 4ac}}{2a}, \tag{8}$$

where

$$a(\tau, s_f) = \frac{\varphi \bar{x}^3}{9\sigma^5 s_f^2}, \qquad b(\tau, s_f) = -\frac{\varphi \bar{x}^2}{3\sigma^3 s_f} + \frac{2\varphi v \bar{x}^3}{9\sigma^5 s_f}; \tag{9a}$$

$$c(\tau, s_f) = -Q + \frac{\varphi \bar{x}}{\sigma} - \frac{\varphi v \bar{x}^2}{3\sigma^3} + \frac{\varphi v \bar{x}^3}{9\sigma^5} + \frac{r\varphi \bar{x}^3}{12\sigma^3}. \tag{9b}$$

Because of the $\bar{x}^3$ associated with $a(\bar{x}, \tau, s_f)$, we observed that the order of accuracy in space might not be at least, $O(\bar{x}^3)$ at the boundary when computing the optimal exercise boundary from its derivative. It is well known that for a stable scheme, a third-order boundary condition is consistent with a fourth-order interior scheme (Adam, 1977). To achieve high order accuracy in space, we present an improvement to

(8) through a method of extrapolation to increase the truncation error of $Q(\bar{x}, \tau)$, up to sixth order accuracy by considering the following lemma:

**Lemma**. Assume $Q(x, \tau) \in C^{n+3}[0, L]$, then it holds

$$a_0 Q(0, \tau) + a_1 Q(\bar{x}, \tau) + \cdots + a_n Q(n\bar{x}, \tau) = c_1 \bar{x} Q'(0, \tau) + c_2 \bar{x}^2 Q''(0, \tau) + c_3 \bar{x}^3 Q'''(0, \tau) + O(\bar{x}^{n+3}), \quad (10a)$$

which gives

$$81 Q(\bar{x}, \tau) - \frac{81}{8} Q(2\bar{x}, \tau) + Q(3\bar{x}, \tau)$$
$$= \frac{175}{4} Q(0, \tau) + \frac{255\bar{x}}{4} Q'(0, \tau) + \frac{99\bar{x}^2}{4} Q''(0, \tau) + \frac{9\bar{x}^3}{2} Q^{(3)}(0, \tau) + O(\bar{x}^6). \quad (10b)$$

**Proof**. Applying the Taylor expansion at 0, we obtain

$$Q(\bar{x}, \tau) = Q(0, \tau) + \bar{x} Q'(0, \tau) + \frac{\bar{x}^2}{2} Q''(0, \tau) + \frac{\bar{x}^3}{6} Q^{(3)}(0, \tau) + \frac{\bar{x}^4}{24} Q^{(4)}(0, \tau) + \frac{\bar{x}^5}{120} Q^{(5)}(0, \tau)$$
$$+ O(\bar{x}^6), \quad (11a)$$

$$Q(2\bar{x}, \tau) = Q(0, \tau) + 2\bar{x} Q'(0, \tau) + 2\bar{x}^2 Q''(0, \tau) + \frac{4\bar{x}^3}{3} Q^{(3)}(0, \tau) + \frac{2\bar{x}^4}{3} Q^{(4)}(0, \tau) + \frac{4\bar{x}^5}{15} Q^{(5)}(0, \tau)$$
$$+ O(\bar{x}^6), \quad (11b)$$

$$Q(3\bar{x}, \tau) = Q(0, \tau) + 3\bar{x} Q'(0, \tau) + \frac{9\bar{x}^2}{2} Q''(0, \tau) + \frac{9\bar{x}^3}{2} Q^{(3)}(0, \tau) + \frac{27\bar{x}^4}{8} Q^{(4)}(0, \tau) + \frac{81\bar{x}^5}{40} Q^{(5)}(0, \tau)$$
$$+ O(\bar{x}^6). \quad (11c)$$

Multiplying (11a) by 16 and subtracting from (11b), we obtain

$$16 Q(\bar{x}, \tau) - Q(2\bar{x}, \tau)$$
$$= 15 Q(0, \tau) + 14 \bar{x} Q'(0, \tau) + 6 \bar{x}^2 Q''(0, \tau) + \frac{4\bar{x}^3}{3} Q^{(3)}(0, \tau) - \frac{2\bar{x}^5}{15} Q^{(5)}(0, \tau)$$
$$+ O(\bar{x}^6). \quad (11d)$$

Multiplying (11b) by 81/16 and subtracting from (11c), we obtain

$$\frac{81}{16} Q(2\bar{x}, \tau) - Q(3\bar{x}, \tau)$$
$$= \frac{65}{16} Q(0, \tau) + \frac{57\bar{x}}{8} Q'(0, \tau) + \frac{45\bar{x}^2}{8} Q''(0, \tau) + \frac{9\bar{x}^3}{4} Q^{(3)}(0, \tau) - \frac{27\bar{x}^5}{40} Q^{(5)}(0, \tau)$$
$$+ O(\bar{x}^6). \quad (11e)$$

Multiplying (11d) by 81/16 and subtracting from (11e), we obtain (10b). Substituting (7) into (10b), we have

$$81Q(\bar{x},\tau) - \frac{81}{8}Q(2\bar{x},\tau) + Q(3\bar{x},\tau) = \frac{255\varphi\bar{x}}{4\sigma} - \frac{33\xi_\tau\varphi\bar{x}^2}{2\sigma^3} + \frac{3\xi_\tau^2\varphi\bar{x}^3}{\sigma^5} + \frac{9r\varphi\bar{x}^3}{4\sigma^3} + O(\bar{x}^6). \quad (11f)$$

The new quadratic form of the derivative of the optimal exercise boundary is then given from (11f) as follows:

$$\left(\frac{\partial s_f}{\partial \tau}\right)^2 + ds_f \frac{\partial s_f}{\partial \tau} + es_f^2 = 0, \quad (12a)$$

where

$$\frac{\partial s_f}{\partial \tau} = \left(\frac{-d - \sqrt{d^2 - 4e}}{2}\right)s_f, \quad d = \left(2 - \frac{11\sigma^2}{2\bar{x}}\right), \quad (12b)$$

$$e(\tau, s_f) = \frac{3\sigma^5}{2\varphi\bar{x}^3}\left[-81Q(\bar{x},\tau) + \frac{81}{8}Q(2\bar{x},\tau) - Q(3\bar{x},\tau) + \left(\frac{255\varphi\bar{x}}{4\sigma} - \frac{33\varphi v\bar{x}^2}{2\sigma^3} + \frac{3\varphi v^2\bar{x}^3}{\sigma^5} + \frac{9r\varphi\bar{x}^3}{4\sigma^3}\right)\right] \quad (12c)$$

With (12), we can ensure at least, a third-order approximation in space at the boundary. To approximate the optimal exercise boundary with high order accuracy in time, we then implement an adaptive fourth-order Runge-Kutta-Fehlberg method for temporal discretization which is detailed in the following section.

## 3. Numerical method

We solve the discretized system of PDEs that consists of the asset, delta, gamma, and speed options in a uniform space grid and non-uniform adaptive time grid $[0, \infty) \times [0\ T]$. We replace the infinite domain with an estimated boundary $x_M$ (Egorova et al., 2016; Kangro and Nicolaides, 2000; Toivanen, 2010,). Let $i$ represent the node points in the grid and M represent the numbers of grid points, respectively, then we have

$$x_i = ih, \quad h = \frac{x_M}{M}, \quad i \in [0, M], \quad (13)$$

Here, the numerical solutions of the asset options, option Greeks, and optimal exercise boundary are represented as $(u)_i^n$, $(w)_i^n$, $(y)_i^n$, $(z)_i^n$, and $s_f^n$.

### 3.1. Fourth-Order Compact Finite Difference Scheme

We employ a compact finite difference scheme (Zhang and Wang, 2012; Bhatt and Khaliq, 2015) for the spatial discretization of our model. For the interior points, we use the compact scheme discretization as follows:

$$f''(x_{i-1}) + 10f''(x_i) + f''(x_{i+1}) = \frac{12}{h^2}[f(x_{i-1}) - 2f(x_i) + f(x_{i+1})] + O(h^4). \quad (14a)$$

For $i = 1$ and $i = M - 1$, we employ a one-sided formula as follows:

$$14f''(x_1) - 5f''(x_2) + 4f''(x_3) - f''(x_4) = \frac{12}{h^2}[f(x_0) - 2f(x_1) + f(x_2)] + O(h^4). \tag{14b}$$

$$14f''(x_{M-1}) - 5f''(x_{M-2}) + 4f''(x_{M-3}) - f''(x_{M-4})$$
$$= \frac{12}{h^2}[f(x_{M-2}) - 2f(x_{M-1}) + f(x_M)] + O(h^4). \tag{14c}$$

The matrix-vector form is as follows:

$$A = \frac{12}{h^2}\begin{bmatrix} -2 & 1 & 0 & \cdots & & & & 0 \\ 1 & -2 & 1 & & & & & \vdots \\ & 1 & -2 & 1 & & & & \\ & & 1 & -2 & 1 & & & \\ 0 & & & \ddots & \ddots & \ddots & & 0 \\ & & & & 1 & -2 & 1 & \\ \vdots & & & & & 1 & -2 & 1 \\ 0 & & & & \cdots & 0 & 1 & -2 \end{bmatrix}_{M-1 \times M-1},$$

$$B = \begin{bmatrix} 14 & -5 & 4 & -1 & 0 & \cdots & 0 \\ 1 & 10 & 1 & & & & \vdots \\ & 1 & 10 & 1 & & & \\ & & 1 & 10 & 1 & & \\ 0 & & & \ddots & \ddots & \ddots & 0 \\ \vdots & & & & 1 & 10 & 1 \\ 0 & \cdots & 0 & -1 & 4 & -5 & 14 \end{bmatrix}_{M-1 \times M-1}, \quad f_u = \frac{12}{h^2}\begin{bmatrix} u_0 \\ 0 \\ 0 \\ \vdots \\ 0 \\ u_M = 0 \end{bmatrix}_{M-1 \times 1};$$

$$f_w = \frac{12}{h^2}\begin{bmatrix} w_0 \\ 0 \\ 0 \\ \vdots \\ 0 \\ w_M = 0 \end{bmatrix}_{M-1 \times 1}, \quad f_y = \frac{12}{h^2}\begin{bmatrix} y_0 \\ 0 \\ 0 \\ \vdots \\ 0 \\ y_M = 0 \end{bmatrix}_{M-1 \times 1}, \quad f_z = \frac{12}{h^2}\begin{bmatrix} z_0 \\ 0 \\ 0 \\ \vdots \\ 0 \\ z_M = 0 \end{bmatrix}_{M-1 \times 1}. \tag{14d}$$

Hence,

$$\mathbf{u}'' = B^{-1}(A\mathbf{u} + \mathbf{f}_u), \qquad \mathbf{w}'' = B^{-1}(A\mathbf{w} + \mathbf{f}_w), \tag{15a}$$

$$\mathbf{y}'' = B^{-1}(A\mathbf{y} + \mathbf{f}_y), \qquad \mathbf{z}'' = B^{-1}(A\mathbf{z} + \mathbf{f}_z). \tag{15b}$$

Substituting (14) into (4), we recast our partial differential equations in the form of a system of ordinary differential equations as follows:

$$\frac{\partial \mathbf{u}}{\partial \tau} = \mathbf{g}_1(\mathbf{u}, \mathbf{w}), \quad \frac{\partial \mathbf{w}}{\partial \tau} = \mathbf{g}_2(\mathbf{w}, \mathbf{u}), \quad \frac{\partial \mathbf{y}}{\partial \tau} = \mathbf{g}_2(\mathbf{y}, \mathbf{w}), \quad \frac{\partial \mathbf{z}}{\partial \tau} = \mathbf{g}_2(\mathbf{z}, \mathbf{y}), \tag{16}$$

where

$$\mathbf{g}_1(\mathbf{u}, \mathbf{w}) = \frac{\sigma^2}{2} B^{-1}(A\mathbf{u} + \mathbf{f}_u) + \xi_\tau \mathbf{w} - r\mathbf{u}, \tag{17a}$$

$$g_2(w, u) = \frac{\sigma^2}{2} B^{-1}(Aw + f_w) + \xi_\tau B^{-1}(Au + f_u) - rw, \tag{17b}$$

$$g_2(y, w) = \frac{\sigma^2}{2} B^{-1}(Ay + f_y) + \xi_\tau B^{-1}(Aw + f_w) - ry, \tag{17c}$$

$$g_2(z, y) = \frac{\sigma^2}{2} B^{-1}(Az + f_z) + \xi_\tau B^{-1}(Ay + f_y) - rz. \tag{17d}$$

We would like to point out some flexibility in this work based on the explicit approach. The rate of change of the optimal exercise boundary is independent of the higher derivatives (delta, gamma, and speed options). By computing the optimal exercise boundary first, we could implement a Dirichlet boundary condition based on (4f). Moreover, the numerical solutions of the asset and delta options with optimal exercise boundary as a coupled system are independent of the higher derivatives (gamma and speed option). The choice of including the gamma and speed options in the coupled system is to approximate them with high order accuracy. Furthermore, it is important to mention that if the choice is to obtain the numerical solutions of the asset option and optimal exercise boundary only, we can further introduce a compact discretization of the first derivative to accommodate such possibility as follows:

$$f'(x_{i-1}) + 4f'(x_i) + f'(x_{i+1}) = \frac{3}{h}[f(x_{i+1}) - f(x_{i-1})] + O(h^4). \tag{18a}$$

For $i = 1$ and $i = M - 1$, we employ a one-sided formula as follows (Zhang and Wang, 2012; Bhatt and Khaliq, 2015):

$$4f'(x_0) + f'(x_1) = \frac{1}{h}\left[-\frac{11}{12}f(x_0) - 4f(x_1) + 6f(x_2) - \frac{4}{3}f(x_3) + \frac{1}{4}f(x_4)\right] + O(h^4). \tag{18b}$$

$$4f'(x_{M-1}) + f'(x_{M-2})$$
$$= \frac{1}{h}\left[\frac{11}{12}f(x_M) - 4f(x_{M-1}) + 6f(x_{M-2}) - \frac{4}{3}f(x_{M-3}) - \frac{1}{4}f(x_{M-4})\right] + O(h^4). \tag{18c}$$

The matrix-vector form is as follows:

$$D = \begin{bmatrix} 4 & 1 & 0 & \cdots & & 0 \\ 1 & 4 & 1 & & & \vdots \\ & 1 & 4 & 1 & & \\ & & 1 & 4 & 1 & \\ 0 & & & \ddots & \ddots & \ddots & 0 \\ \vdots & & & & 1 & 4 & 1 \\ 0 & & & \cdots & 0 & 1 & 4 \end{bmatrix}_{M-1 \times M-1}, \quad \bar{f}_u = \frac{11}{12h}\begin{bmatrix} -u_0 \\ 0 \\ 0 \\ \vdots \\ 0 \\ u_M = 0 \end{bmatrix}_{M-1 \times 1}, \tag{18d}$$

$$C = \frac{3}{h}\begin{bmatrix} -\frac{4}{3} & 2 & -\frac{4}{9} & \frac{1}{12} & 0 & \cdots & & & 0 \\ -1 & 0 & 1 & & & & & & \vdots \\ & -1 & 0 & 1 & & & & & \\ & & -1 & 0 & 1 & & & & \\ 0 & & & \ddots & \ddots & \ddots & & & 0 \\ & & & & -1 & 0 & 1 & & \\ \vdots & & & & & -1 & 0 & 1 & \\ 0 & & & \cdots & & -\frac{1}{12} & \frac{4}{9} & -2 & \frac{4}{3} \end{bmatrix}_{M-1 \times M-1}, \qquad (18e)$$

where $\boldsymbol{u}' = D^{-1}(C\boldsymbol{u} + \bar{\boldsymbol{f}}_u)$. Implementing it in (3), we then have

$$\frac{\partial \boldsymbol{u}}{\partial \tau} = \frac{\sigma^2}{2} B^{-1}(A\boldsymbol{u} + \boldsymbol{f}_u) + \xi_\tau D^{-1}(C\boldsymbol{u} + \bar{\boldsymbol{f}}_u) - r\boldsymbol{u}, \qquad (18f)$$

which can be used to obtain the numerical solutions of the optimal exercise boundary and the asset option.

### 3.2. Adaptive and Classical Fourth-Order Time Integrators

**Adaptive Runge-Kutta-Fehlberg Method:** By recasting our system of discretized partial differential equations in the form of ordinary differential equations, we then present (16)-(17) in explicit form as follows:

$$\frac{\partial \boldsymbol{u}^n}{\partial \tau} = \frac{\sigma^2}{2} B^{-1}(A\boldsymbol{u}^n + \boldsymbol{f}_u^n) + \xi_n \boldsymbol{w}^n - r\boldsymbol{u}^n, \qquad (19a)$$

$$\frac{\partial \boldsymbol{w}^n}{\partial \tau} = \frac{\sigma^2}{2} B^{-1}(A\boldsymbol{w}^n + \boldsymbol{f}_w^n) + \xi_n B^{-1}(A\boldsymbol{u}^n + \boldsymbol{f}_u^n) - r\boldsymbol{w}^n, \qquad (19b)$$

$$\frac{\partial \boldsymbol{y}^n}{\partial \tau} = \frac{\sigma^2}{2} B^{-1}(A\boldsymbol{y}^n + \boldsymbol{f}_y^n) + \xi_n B^{-1}(A\boldsymbol{w}^n + \boldsymbol{f}_w^n) - r\boldsymbol{y}^n, \qquad (19c)$$

$$\frac{\partial \boldsymbol{z}^n}{\partial \tau} = \frac{\sigma^2}{2} B^{-1}(A\boldsymbol{z}^n + \boldsymbol{f}_z^n) + \xi_n B^{-1}(A\boldsymbol{y}^n + \boldsymbol{f}_y^n) - r\boldsymbol{z}^n. \qquad (19d)$$

In this work, we implement a fourth-order adaptive Runge-Kutta-Fehlberg method (Fehlberg, 1969) based on the coefficients of Cash and Karp (1990). Runge-Kutta-Fehlberg method uses a fifth-order Runge-Kutta method to estimate the local truncation error of the fourth-order Runge-Kutta method (Burden et al., 2016). With a given tolerance, the optimal time step is obtained for each time level. For brevity, we only describe function which follows from (19) for computing the new values and error of the asset option from the RKF as follows:

The fourth-order and fifth-order Runge-Kutta methods are given as

$$u^{n+1} = u^n + \left(\frac{37}{378}L_u^1 + \frac{250}{621}L_u^3 + \frac{125}{594}L_u^4 + \frac{512}{1771}L_u^6\right), \tag{20a}$$

$$\bar{u}^{n+1} = u^n + \left(\frac{2825}{27648}L_u^1 + \frac{18575}{48384}L_u^3 + \frac{13525}{55296}L_u^4 + \frac{277}{14336}L_u^5 + \frac{1}{4}L_u^6\right), \tag{20b}$$

respectively, and the error estimated as

$$e_u = \|\bar{u}^{n+1} - u^{n+1}\|_\infty < \varepsilon, \tag{20c}$$

where

$$L_u^1 = g_1(u^n, w^n)k, \qquad L_u^2 = g_1\left(u^n + \frac{1}{5}L_u^1, w^n + \frac{1}{5}L_w^1\right)k; \tag{20d}$$

$$L_u^3 = g_1\left(u^n + \frac{3}{40}L_u^1 + \frac{9}{40}L_u^2, w^n + \frac{3}{40}L_w^1 + \frac{9}{40}L_w^2\right)k, \tag{20e}$$

$$L_u^4 = g_1\left(u^n + \frac{3}{10}L_u^1 - \frac{9}{10}L_u^2 + \frac{6}{5}L_u^3, w^n + \frac{3}{10}L_w^1 - \frac{9}{10}L_w^2 + \frac{6}{5}L_w^3\right)k, \tag{20f}$$

$$L_u^5 = g_1\left(u^n - \frac{11}{54}L_u^1 + \frac{5}{2}L_u^2 - \frac{70}{27}L_u^3 + \frac{35}{27}L_u^4, w^n - \frac{11}{54}L_w^1 + \frac{5}{2}L_w^2 - \frac{70}{27}L_w^3 + \frac{35}{27}L_w^4\right)k, \tag{20g}$$

$$L_u^6 = g_1\left(u^n + \frac{1631}{55296}L_u^1 + \frac{175}{512}L_u^2 + \frac{575}{13824}L_u^3 + \frac{44275}{110592}L_u^4 + \frac{253}{4096}L_u^5, w^n + \frac{1631}{55296}L_w^1 + \frac{175}{512}L_w^2 \right.$$
$$\left. + \frac{575}{13824}L_w^3 + \frac{44275}{110592}L_w^4 + \frac{253}{4096}L_w^5\right)k. \tag{20h}$$

Here, $k$ represents the time step. Similarly, the mathematical formulation in (20) also follows for computing the option Greeks. Hence, for the sake of brevity, we skip them. If the condition in (20c) fails based on an arbitrary $\varepsilon$, an optimal parameter is determined, from which a new time step is calculated until an optimal time step that satisfied (20c) is obtained. Moreover, if the condition in (20c) is satisfied, a new time step is also estimated which will be used in the next time level. The calculation is done (William and Saul, 1992; Clayton et al., 2019) as follows:

$$k_{new} = \begin{cases} 0.9 k_{old}(Tol/e_u)^{1/5}, & \varepsilon \leq e_u, \\ 0.9 k_{old}(Tol/e_u)^{1/4}, & \varepsilon > e_u. \end{cases} \tag{21}$$

**Classical Runge-Kutta Method:** Here, as a comparison and to calculate the convergent rate with a constant time step, we employ a fourth-order explicit Runge-Kutta method (RK4) for temporal discretization. We fully describe the procedure for solving (19) using the Runge-Kutta method as follows:

$$R_u^1 = g_1(u^n, w^n)k, \qquad R_w^1 = g_2(w^n, u^n)k; \tag{22a}$$

$$R_y^1 = g_2(y^n, w^n)k, \qquad R_z^1 = g_2(z^n, y^n)k; \tag{22b}$$

$$R_u^2 = g_1\left(u^n + \frac{1}{2}R_u^1, w^n + \frac{1}{2}R_w^1\right)k, \qquad R_w^2 = g_2\left(w^n + \frac{1}{2}R_w^1, u^n + \frac{1}{2}R_u^1\right)k; \tag{22c}$$

$$R_y^2 = g_2\left(y^n + \frac{1}{2}R_y^1, w^n + \frac{1}{2}R_w^1\right)k, \qquad R_z^2 = g_2\left(z^n + \frac{1}{2}R_z^1, y^n + \frac{1}{2}R_y^1\right)k; \tag{22d}$$

$$R_u^3 = g_1\left(u^n + \frac{1}{2}R_u^2, w^n + \frac{1}{2}R_w^2\right)k, \qquad R_w^3 = g_2\left(w^n + \frac{1}{2}R_w^2, u^n + \frac{1}{2}R_u^2\right)k; \tag{22e}$$

$$R_y^3 = g_2\left(y^n + \frac{1}{2}R_y^2, w^n + \frac{1}{2}R_w^2\right)k, \qquad R_z^3 = g_2\left(z^n + \frac{1}{2}R_z^2, y^n + \frac{1}{2}R_y^2\right)k; \tag{22f}$$

$$R_u^4 = g_1(u^n + R_u^3, w^n + R_w^3)k, \qquad R_w^4 = g_2(w^n + R_w^3, u^n + R_u^3)k; \tag{22g}$$

$$R_y^4 = g_2(y^n + R_y^3, w^n + R_w^3)k, \qquad R_z^4 = g_2(z^n + R_z^3, y^n + R_y^3)k; \tag{22h}$$

$$u^{n+1} = u^n + \frac{k}{6}(R_u^1 + 2R_u^2 + 2R_u^3 + R_u^4), \qquad w^{n+1} = w^n + \frac{k}{6}(R_w^1 + 2R_w^2 + 2R_w^3 + R_w^4); \tag{23a}$$

$$y^{n+1} = y^n + \frac{k}{6}(R_y^1 + 2R_y^2 + 2R_y^3 + R_y^4), \qquad z^{n+1} = z^n + \frac{k}{6}(R_z^1 + 2R_z^2 + 2R_z^3 + R_z^4). \tag{23b}$$

**Approximation of the Optimal Exercise Boundary**: Because of the explicit nature of our proposed method, we need to approximate the optimal exercise boundary before computing the asset option and option Greeks. To achieve this, we discretize (12b) using both adaptive and classical RK4 methods.

Let

$$\frac{\partial s_f^n}{\partial \tau} = g_3(s_f^n, u_{\bar{x}}^n) = \left(\frac{-d - \sqrt{d^2 - 4e^n}}{2}\right) s_f^n, \tag{24a}$$

with

$$e^n = \frac{3\sigma^5}{2\varphi\bar{x}^3}\left[-81Q_{\bar{x}}^n + \frac{81}{8}Q_{2\bar{x}}^n - Q_{3\bar{x}}^n + \left(\frac{255\varphi\bar{x}}{4\sigma} - \frac{33\varphi v \bar{x}^2}{2\sigma^3} + \frac{3\varphi v^2 \bar{x}^3}{\sigma^5} + \frac{9r\varphi\bar{x}^3}{4\sigma^3}\right)\right] \tag{24c}$$

For the adaptive Runge-Kutta-Fehlberg method, the fourth order Runge-Kutta method

$$s_f^{n+1} = s_f^n + \left(\frac{37}{378}R_{s_f}^1 + \frac{250}{621}R_{s_f}^3 + \frac{125}{594}R_{s_f}^4 + \frac{512}{1771}R_{s_f}^5\right), \tag{25a}$$

is computed simultaneously with the fifth-order Runge-Kutta method

$$\bar{s}_f^{n+1} = s_f^n + \left(\frac{2825}{27648}R_{s_f}^1 + \frac{18575}{48384}R_{s_f}^3 + \frac{13525}{55296}R_{s_f}^4 + \frac{277}{14336}R_{s_f}^5 + \frac{1}{4}R_{s_f}^6\right), \tag{25b}$$

and the error estimated as

$$e_{s_f} = |s_f^{n+1} - \bar{s}_f^{n+1}| < \varepsilon, \qquad (25c)$$

where

$$R_{s_f}^1 = g_3(s_f^n, u_{\bar{x}}^n)k, \qquad R_{s_f}^2 = g_3\left(s_f^n + \frac{1}{5}R_{s_f}^1, u_{\bar{x}}^n\right)k; \qquad (25d)$$

$$R_{s_f}^3 = g_3\left(s_f^n + \frac{3}{40}R_{s_f}^1 + \frac{9}{40}R_{s_f}^2, u_{\bar{x}}^n\right)k, \qquad (25e)$$

$$R_{s_f}^4 = g_3\left(s_f^n + \frac{3}{10}R_{s_f}^1 - \frac{9}{10}R_{s_f}^2 + \frac{6}{5}R_{s_f}^3, u_{\bar{x}}^n\right)k, \qquad (25f)$$

$$R_{s_f}^5 = g_3\left(s_f^n - \frac{11}{54}R_{s_f}^1 + \frac{5}{2}R_{s_f}^2 - \frac{70}{27}R_{s_f}^3 + \frac{35}{27}R_{s_f}^4, u_{\bar{x}}^n\right)k, \qquad (25g)$$

$$R_{s_f}^6 = g_3\left(s_f^n + \frac{1631}{55296}R_{s_f}^1 + \frac{175}{512}R_{s_f}^2 + \frac{575}{13824}R_{s_f}^3 + \frac{44275}{110592}R_{s_f}^4 + \frac{253}{4096}R_{s_f}^5, u_{\bar{x}}^n\right)k. \qquad (25h)$$

For the classical Runge-Kutta method, we compute as follows:

$$R_{s_f}^1 = g_3(s_f^n, u_{\bar{x}}^n)k, \qquad R_{s_f}^2 = g_3\left(s_f^n + \frac{1}{2}R_{s_f}^1, u_{\bar{x}}^n\right)k, \qquad R_{s_f}^3 = g_3\left(s_f^n + \frac{1}{2}R_{s_f}^2, u_{\bar{x}}^n\right)k; \qquad (26a)$$

$$R_{s_f}^4 = g_3\left(s_f^n + R_{s_f}^3, u_{\bar{x}}^n\right)k, \qquad s_f^{n+1} = s_f^n + \frac{1}{6}\left(R_{s_f}^1 + 2R_{s_f}^2 + 2R_{s_f}^3 + R_{s_f}^4\right). \qquad (26b)$$

Here, we choose $\bar{x} = 2h$ in our numerical experiment.

### 3.3. Computational Procedure using Adaptive RKF Time Integrator

In this section, we describe the implementation and algorithm for computing the asset, delta, gamma, and speed options using the adaptive Runge-Kutta-Fehlberg methods based on the coefficients of Cash and Karp (1990). It is worth noting that in this work, we restrict the error estimate only with the asset option. That is, we use only $e_u$ to confirm to optimal time step.

When approximating our numerical solutions, there is a threshold for $k$ above which the optimal exercise boundary, when computed from the quadratic equation, will give a complex value. We first adapt our code to find a maximum $k$ that guarantees a real value for the optimal exercise boundary before proceeding to find and obtain the optimal time step and numerical approximation(s) at each time level, respectively. Algorithms for obtaining the numerical solutions of the optimal exercise boundary, asset option, and the option Greeks using the fourth-order adaptive Runge-Kutta methods are described below.

**Algorithm.** Algorithm for the Runge-Kutta-Fehlberg method (RKF).

1. initialize $t = 0, h, k, T, A, B,$ and $Tol$ ▷ The initial choice of $k$ is arbitrary and independent of $h$
2. initialize $s_f^n, \boldsymbol{u}^n, \boldsymbol{w}^n, \boldsymbol{y}^n$ and $\boldsymbol{z}^n$
3. **while** $t < T$
4.    **if** $t + k > T$
5.       $k = T - t$
6.    **endif**
7.    **while** true
8.       compute $s_f^{n+1}$ ▷ based on (25)
9.       **if** $s_f^{n+1}$ is a real value, break ▷ obtain a maximum $k$ that guarantee real value for $s_f^{n+1}$
10.      **else** $k = \phi k$ ▷ $0.1 \leq \phi \leq 0.5$ if the initial choice of $k = h$
11.      **endif**
12.    **end while**
13.    compute $\boldsymbol{f}_u, \boldsymbol{f}_w, \boldsymbol{f}_y,$ and $\boldsymbol{f}_z$
14.    compute $\boldsymbol{u}^{n+1}, \bar{\boldsymbol{u}}^{n+1}, \boldsymbol{w}^{n+1}, \boldsymbol{y}^{n+1},$ and $\boldsymbol{z}^{n+1}$ ▷ based on (22) and (23)
15.    compute $e_u = \|\bar{\boldsymbol{u}}^{n+1} - \boldsymbol{u}^{n+1}\|_\infty$,
16.    **if** $e_u < Tol$
17.       set $\boldsymbol{u}^n = \boldsymbol{u}^{n+1}, \boldsymbol{w}^n = \boldsymbol{w}^{n+1}, \boldsymbol{y}^n = \boldsymbol{y}^{n+1}, \boldsymbol{z}^n = \boldsymbol{z}^{n+1}$ and $s_f^n = s_f^{n+1}$
18.       set $\delta_u = 0.9(Tol/e_u)^{1/4}$ and $k = \delta_u k$ ▷ based on (24)
19.       $t = t + k$
20.    **else**
21.       set $\delta_u = 0.9(Tol/e_u)^{1/5}$ and $k = \delta_u k$ ▷ based on (24)
22.    **endif**
23. **repeat**

## 4. Numerical Experiment and Discussion

In this section, the numerical performance of the proposed method is investigated and validated using two examples and further compared with the existing results. The numerical experiment was carried out on the mesh with a uniform grid size.

**Example 1.** Consider the example provided in the work of Zhu (2006). The following data are presented

$$K = 100, \quad T = 1, \quad r = 10\%, \quad \sigma = 30\%. \tag{27}$$

In this example, we focus on comparing the values of the optimal exercise boundary. We compared the results of the adaptive Runge-Kutta-Fehlberg method (RKF) coupled with a finite compact scheme (FCS-RKF) with that of the method of Zhu (2006), the numerical method of Wu and Kwok (1997), and the classical Runge-Kutta method with a finite compact scheme which we label as FCS-RK4.

The results were listed in Table 1. The plots of the asset option, option Greeks, and optimal exercise boundary were displayed in Figs. 1. From Table 1, one can observe that our numerical approximations of the optimal exercise boundary for the FCS-RK4 and FCS-RKF are very close to the analytical approximation of Zhu (2006). Furthermore, we further observe from Fig. 1 that the FCS-RK4 method does not approximate the higher derivative (i.e., speed option) accurately when compared with the FCS-RKF.

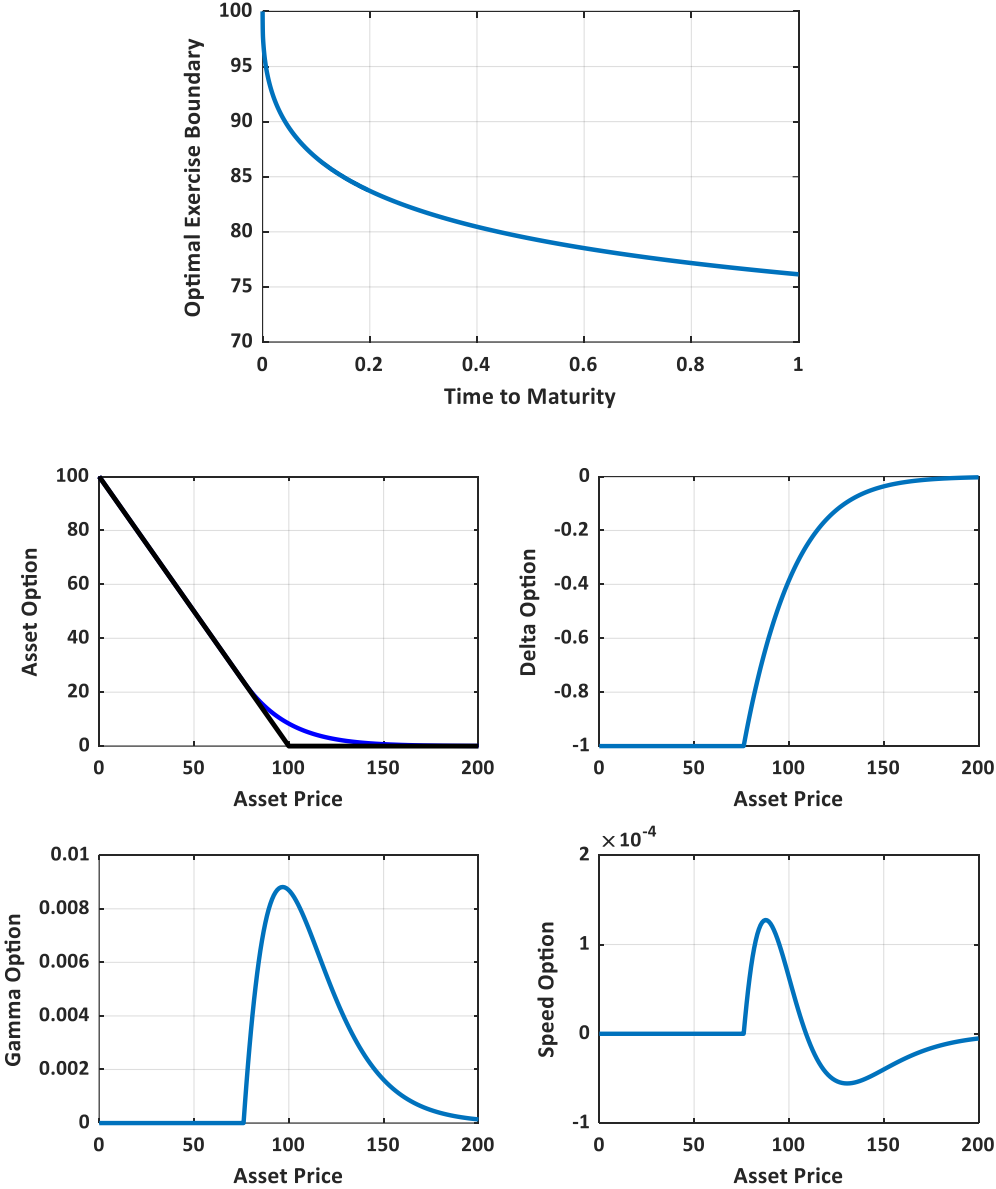

**Fig. 1a**. Asset option, option Greeks and optimal exercise boundary with FCS-RKF ($h = 0.01$, $\tau = T$, $\varepsilon = 10^{-8}$).

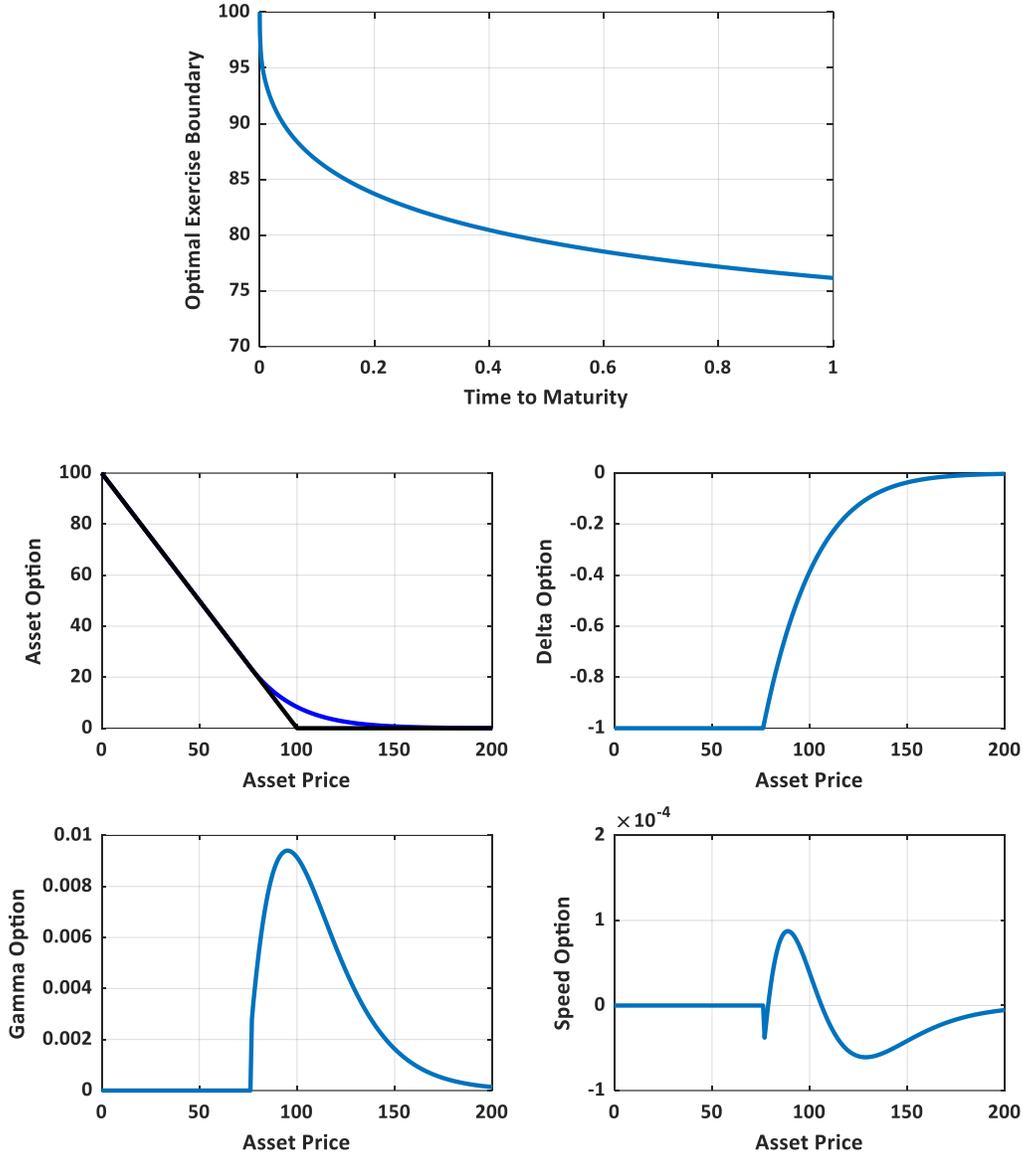

**Fig. 1b**. Asset option, option Greeks, and optimal exercise boundary with FCS-RK4 ($h = 0.01, k = 0.0001$ $\tau = T$).

**Table 1.** Comparison of the optimal exercise boundary for example 1 ($\tau = T$, $h = 0.01$).

| | Optimal Exercise Boundary | | |
|---|---|---|---|
| Zhu (2007) | Wu and Kwok (1997) | FCS-RK4 | FCS-RKF |
| 76.11 | 76.25 | 76.16 | 76.16 |

**Example 2.** Consider the example in the work of Kim et al. (2013). We compare our result with Kim et al. (2013), the moving boundary method (Muthuraman, 2008), and the Binomial method (Cox et al., 1979) which is used as the benchmark result. Here, we use (18) to compute only the option values and the optimal exercise boundary. The following data are considered

$$K = 100, \quad T = 0.5, \quad r = 5\%, \quad \sigma = 20\%. \tag{28}$$

The results were listed in Table 2. In Tables 3 and 4, we present the total CPU time(s), optimal exercise boundary value, and the global minimum and maximum optimal time step of the FCS-RKF based on varying tolerance $\varepsilon$ and step size. The plot of adaptive optimal time step selection for each time level based on varying tolerance $\varepsilon$ and step size was further displayed in Fig. 2.

From Table 2, one can easily see the better performance of the FCS-RKF. With $\varepsilon = 10^{-8}$ and from $h = 0.0125$, the results obtained from the FCS-RKF are very close to the one obtained from the binomial method that serves as a benchmark in this example. From Tables 3 and 4, we observe the dependent of the optimal time selection on the tolerance $\varepsilon$ and fixed step size in direct proportion. From Fig. 2, we observe the concentration of small oscillation near the payoff ($\tau = 0$) as the tolerance and the step size is reduced. This is expected because of the variation due to the discontinuity in the first derivative of the asset option that normally occurs at the payoff. Hence, a small varying time step is needed in the region near the payoff. Furthermore, it is important to observe from Fig. 2 that the adaptive method could be very useful in detecting unknown locations of discontinuity or rapid variation in systems (Gear and Østerby, 1984; Dieci and Lopez, 2012) when it is implemented efficiently.

**Table 2.** Comparison of the asset option in example 2.

| S | Binomial | MBM | Kim et al. | FCS-RK4 |
|---|---|---|---|---|
| 80 | 20.0000 | 20.0000 | 20.0000 | 20.0000 |
| 90 | 10.6661 | 10.6680 | 10.6661 | 10.6680 |
| 100 | 4.6556 | 4.6504 | 4.6549 | 4.6588 |
| 110 | 1.6681 | 1.6629 | 1.6686 | 1.6704 |
| 120 | 0.4976 | 0.4993 | 0.4985 | 0.4988 |
| S | FCS-RKF ($\varepsilon = 10^{-8}$) | | | |
|  | $h = 0.025$ | $h = 0.0125$ | $h = 0.01$ | |
| 80 | 20.0000 | 20.0000 | 20.0000 | |
| 90 | 10.6653 | 10.6660 | 10.6661 | |
| 100 | 4.6539 | 4.6554 | 4.6555 | |
| 110 | 1.6666 | 1.6678 | 1.6679 | |
| 120 | 0.4968 | 0.4974 | 0.4975 | |

**Table 3.** Performance of the FCS-RKF based on fixed step size ($h = 0.01$) and varying tolerance.

| S | $\varepsilon = 10^{-3}$ | $\varepsilon = 10^{-5}$ | $\varepsilon = 10^{-8}$ |
|---|---|---|---|
| CPU time(s) | 23.61 | 57.57 | 1667.53 |
| $s_f(\tau)$ | 83.7835 | 83.9083 | 83.9195 |
| Max. time step | 1.44e-3 | 2.88e-4 | 9.04e-6 |
| Min. time step | 2.85e-4 | 2.18e-5 | 3.45e-6 |

**Table 4.** Performance of the FCS-RKF based on fixed tolerance $\varepsilon = 10^{-8}$ and varying step size.

| S | $h = 0.1$ | $h = 0.05$ | $h = 0.01$ |
|---|---|---|---|
| CPU time(s) | 3.93 | 19.97 | 1667.53 |
| $s_f(\tau)$ | 84.6165 | 83.9416 | 83.9195 |
| Max. time step | 7.90e-4 | 1.55e-4 | 9.04e-6 |
| Min. time step | 3.74e-4 | 8.99e-5 | 3.45e-6 |

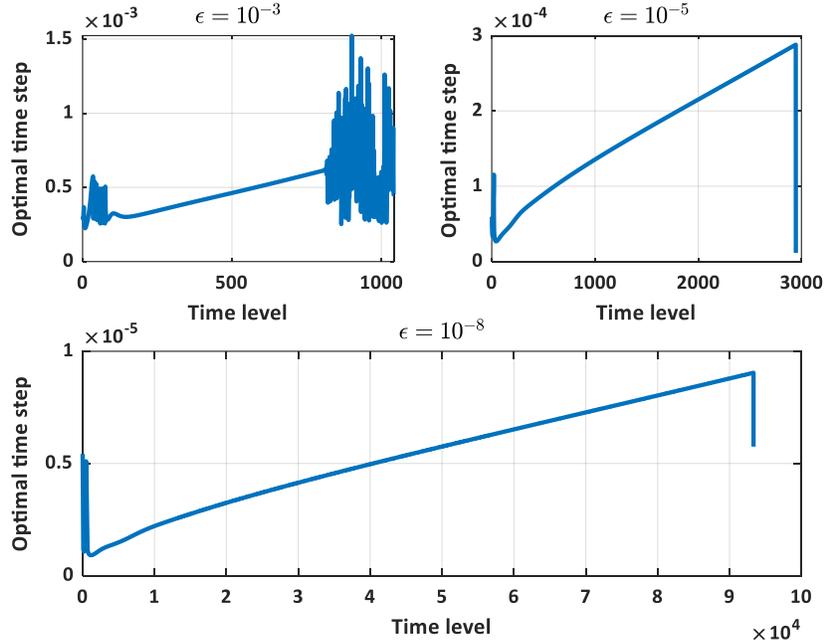

**Fig. 2a.** Optimal time step selection for each time level using FCS-RKF with a fixed $h = 0.01$.

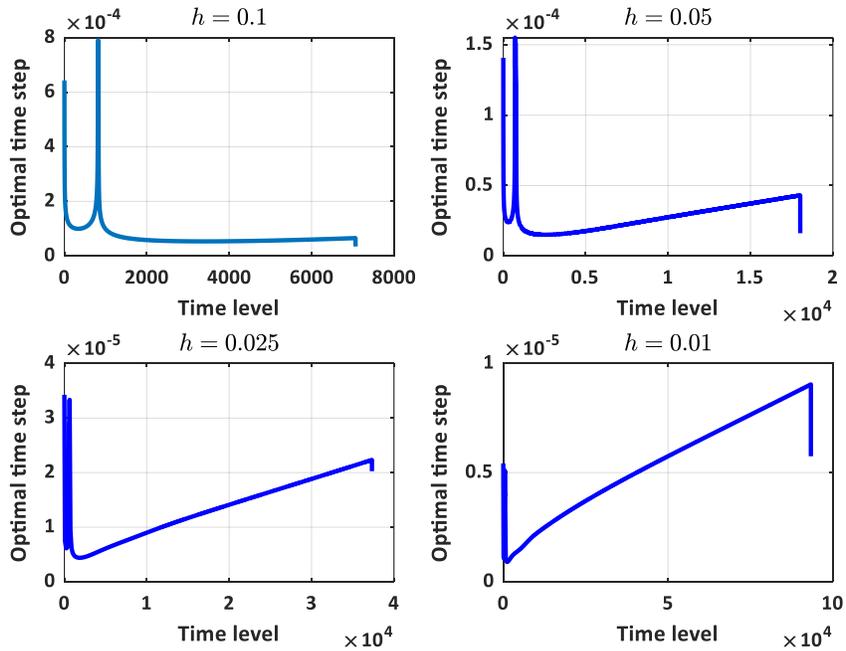

**Fig. 2b.** Optimal time step selection for each time level using FCS-RKF with a fixed $\varepsilon = 10^{-8}$.

To check the maximum errors and convergent rates in space, we used (12), example 1, and FCS-RK4 and selected a constant time step $k = 1 \times 10^{-6}$ with varying step size $h = 0.2, 0.1, 0.05, 0.025, 0.0125$ and 0.00625. We then compared the convergence rate in space based on the extrapolation method we implemented in (12) with the method of Kim et al. (2013) in (8). We further check the maximum errors and the convergent rates of the delta option based on the method of extrapolation with a constant time step $k = 1 \times 10^{-6}$ and varying step size $h = 0.2, 0.1, 0.05, 0.025, 0.0125$ and 0.00625. The results are displayed in Table 5. For the asset option, the method of Kim et al. (2013) provides up to second order in space as seen in Table 5. However, by improving it with a method of extrapolation, we obtained a high order accuracy in space for both the asset and delta options which is in close agreement with the theoretical convergent rate.

**Table 5.** Maximum errors and convergent rates in space of the asset and delta options with $k = 1 \times 10^{-6}$.

| | Asset option | | | |
|---|---|---|---|---|
| | Kim et al. (2013) | | Our method | |
| $h$ | maximum error | convergent rate | maximum error | convergent rate |
| 0.2 | ~ | ~ | ~ | ~ |
| 0.1 | $6.033 \times 10^{-1}$ | ~ | $1.263 \times 10^{-0}$ | ~ |
| 0.05 | $7.147 \times 10^{-1}$ | $-0.236$ | $7.147 \times 10^{-2}$ | 4.142 |
| 0.025 | $1.693 \times 10^{-1}$ | 2.070 | $7.470 \times 10^{-3}$ | 3.258 |
| 0.0125 | $4.100 \times 10^{-2}$ | 2.047 | $5.440 \times 10^{-4}$ | 3.780 |
| 0.000625 | $1.013 \times 10^{-2}$ | 2.015 | $1.331 \times 10^{-5}$ | 5.353 |
| Average CR | | 1.474 | | 4.133 |
| | Delta Option (Our method) | | | |
| | maximum error | | convergent rate | |
| 0.2 | ~ | | ~ | |
| 0.1 | $1.822 \times 10^{-0}$ | | ~ | |
| 0.05 | $3.756 \times 10^{-1}$ | | 2.279 | |
| 0.025 | $2.127 \times 10^{-2}$ | | 4.143 | |
| 0.0125 | $1.099 \times 10^{-3}$ | | 4.274 | |
| 0.00625 | $2.238 \times 10^{-5}$ | | 5.618 | |
| Average CR | | | 4.078 | |

## 5. Conclusion

We have proposed an adaptive and explicit fourth-order Runge-Kutta-Fehlberg method with a fourth-order compact scheme for pricing American options. By implementing logarithmic transformation, taking a further derivative, improving the method of Kim et al. (2013) by adopting a method of extrapolation, we then obtain an analytical formula for the velocity of the optimal exercise boundary with high order accuracy in space. We further recast the free boundary problem to a system of coupled ordinary differential equations, employ compact finite difference for spatial discretization with Dirichlet boundary condition and implement an adaptive fourth-order Runge-Kutta-Fehlberg method for temporal discretization. This enables us to approximate the optimal exercise boundary, options value, and option Greeks in the set of coupled ODEs with high order accuracy both in space and in time. Furthermore, we check the convergent rate of our numerical method with the FCS-RK4 method and confirm that the numerical convergent rate is in close agreement with the theoretical convergent rate. By further comparing the result from the FCS-RKF method with the existing methods including the classical Runge-Kutta (FCS-RK4), we then validate the superiority of the adaptive method.

## Acknowledgment

This is a preprint of an article published in Japan Journal of Industrial and Applied Mathematics (JJIAM). The final authenticated version is available online at: https://doi.org/10.1007/s13160-021-00470-2.